%% file: otoc_v15.tex
\documentclass[aps,pra,superscriptaddress,twocolumn,10pt,longbibliography]{revtex4-1}

\usepackage{graphicx}
\usepackage{amsmath}
\usepackage{amssymb}
\usepackage{dsfont}
\usepackage{xspace}
\usepackage{color}
\usepackage[usenames,dvipsnames]{xcolor}
\usepackage[outercaption]{sidecap}   
\usepackage[colorlinks=true,linkcolor=blue,urlcolor=blue,citecolor=blue]{hyperref}
\usepackage[normalem]{ulem}
\usepackage{bbm}
\usepackage{array}

\DeclareMathOperator{\tr}{\mbox{tr}}
\DeclareMathOperator{\re}{\mbox{Re}}
\DeclareMathOperator{\im}{\mbox{Im}}

\newcommand{\nag}{{\phantom{\dagger}}}

\newcommand{\eqw}[1]{(\ref{#1})}
\newcommand{\eq}[1]{Eq.~(\ref{#1})}

\newcommand{\tab}[1]{Tab.\thinspace{}\ref{#1}}

\newcommand{\fig}[1]{Fig.\thinspace{}\ref{#1}}

\newcommand{\fc}[1]{({#1})}
\newcommand{\figc}[2]{Fig.\thinspace{}\ref{#1}\thinspace{}\fc{#2}}

\def\bra#1{\mathinner{\langle{#1}|}}
\def\ket#1{\mathinner{|{#1}\rangle}}

\usepackage{times}

\begin{document}

\title{Scrambling and thermalization in a diffusive quantum many-body system}

\author{A. Bohrdt}%
\affiliation{Department of Physics, Walter Schottky Institute, and Institute for Advanced Study, Technical University of Munich, 85748 Garching, Germany}%
\affiliation{Department of Physics, Harvard University, Cambridge, MA 02138, USA}%
\author{C. B. Mendl}%
\affiliation{Stanford Institute for Materials and Energy Sciences, SLAC National Accelerator Laboratory, and Stanford University, CA 94025, USA}%
\author{M. Endres}%
\affiliation{Division of Physics, Mathematics and Astronomy, California Institute of Technology, Pasadena, CA 91125, USA}%
\author{M. Knap}%
\affiliation{Department of Physics, Walter Schottky Institute, and Institute for Advanced Study, Technical University of Munich, 85748 Garching, Germany}%

\date{\today}

\begin{abstract}

Out-of-time ordered (OTO) correlation functions describe scrambling of information in correlated quantum matter. They are of particular interest in incoherent quantum systems lacking well defined quasi-particles. Thus far, it is largely elusive how OTO correlators spread in incoherent systems with diffusive transport governed by a few globally conserved quantities. Here, we study the dynamical response of such a system using high-performance matrix-product-operator techniques. Specifically, we consider the non-integrable, one-dimensional Bose-Hubbard model in the incoherent high-temperature regime. Our system exhibits diffusive dynamics in time-ordered correlators of globally conserved quantities, whereas OTO correlators display a ballistic, light-cone spreading of quantum information. The slowest process in the global thermalization of the system is thus diffusive, yet information spreading is not inhibited by such slow dynamics. We furthermore develop an experimentally feasible protocol to overcome some challenges faced by existing proposals and to probe time-ordered and OTO correlation functions. Our study opens new avenues for both the theoretical and experimental exploration of thermalization and information scrambling dynamics.

\end{abstract}

\pacs{
}

\maketitle

Dynamical correlations of many-body quantum systems provide direct information about many-body excitations~\cite{fetter_1971}, describe quantum phases and transitions~\cite{sachdev_quantum_2011}, and characterize certain topological aspects~\cite{punk_topological_2014,morampudi_statistics_2016}. The dynamical response of a many-body system to a local perturbation is obtained from a time ordered correlation function, $\langle \hat W(t) \hat V(0) \rangle$, which describes the \emph{relaxation} of the many-body system following the initial excitation by the operator $\hat V$ that is then probed at later times by $\hat W$. However, in general such time-ordered correlation functions cannot capture the spread of information across a quantum system, especially in a regime where quasiparticles are not well-defined. 

\begin{figure}[b!]
  \centering 
  \includegraphics[width=.46\textwidth]{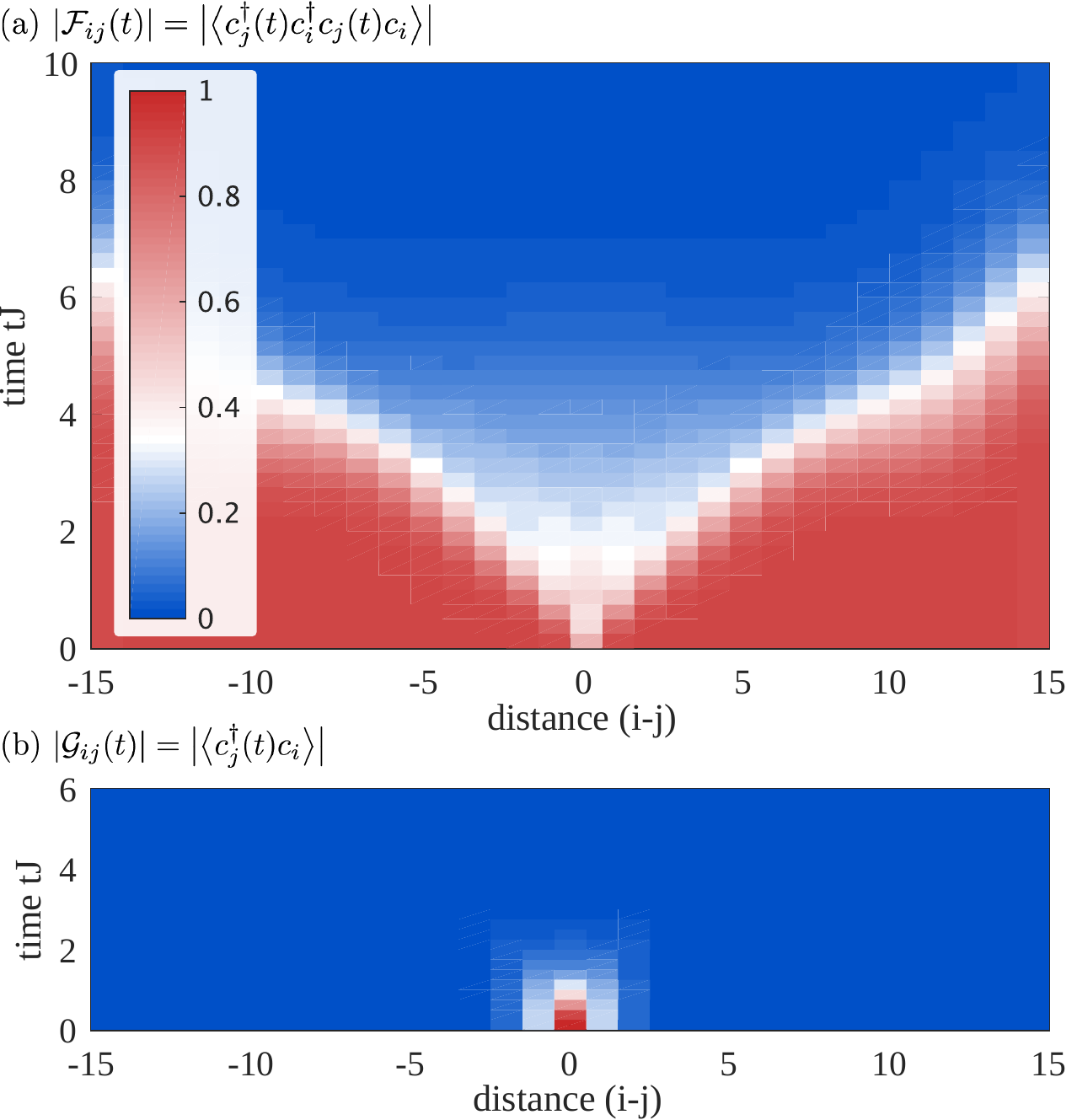}
  \caption{ \textbf{Dynamical correlation functions in the incoherent transport regime.} \fc{a} Out-of-time ordered (OTO) correlators measure the scrambling of information across a quantum state. We compute OTO correlators $\mathcal{F}_{ij}(t)=\langle c_j^\dag(t) c_i^\dag c_j(t) c_i \rangle$ in the 1D Bose-Hubbard model at high temperature $T=4J$ for interactions $U=J$, chemical potential $\mu=0$, and system size $L=30$. In the high temperature regime, well-defined quasiparticles cease to exist. However, the OTO correlator $\mathcal{F}_{ij}$ exhibits a light-cone spreading of information.  
  \fc{b} The breakdown of well-defined quasiparticles is demonstrated by the one-particle Green's function $\mathcal{G}_{ij}(t)=\langle c_j^\dag(t) c_i \rangle$, which quickly decays to zero within $\tau J \sim 0.6$. The lifetime is thus shorter than the hopping rate, indicating a regime of incoherent transport.
  } 
  \label{fig:otoc}
\end{figure}

Recently, it has been proposed that spreading or ``\emph{scrambling}'' of quantum information across all the system's degrees of freedom can be characterized by out-of-time ordered (OTO) correlation functions: $\langle \hat W^\dag(t) \hat V^\dag(0) \hat W(t) \hat V(0) \rangle$~\cite{larkin_1969, shenker_black_2014, kitaev_2014, roberts_localized_2015, maldacena_bound_2016, hosur_chaos_2016}.
These correlation functions  appear as the out-of-time ordered part of $\langle |[\hat W(t),\hat V(0)]|^2 \rangle$ and hence predict the growth of the squared commutator between $\hat W(t)$ and $\hat V(0)$. OTO correlators are thus capable of describing a quantum analogue of the butterfly effect in classical chaotic systems, which characterizes the spread of local excitations over the whole system. At short times, OTO correlators are expected to grow exponentially with a rate characterized by the Lyapunov exponent $\lambda_L$. The Lyapunov exponent has been conjectured to be bounded by $0 \leq \lambda_L \leq 2\pi T$~\cite{maldacena_bound_2016}. This bound is saturated in strongly coupled field theories with a gravity dual~\cite{shenker_black_2014} and in disordered models describing a strange metal~\cite{sachdev_gapless_1993, kitaev_2014, banerjee_solvable_2016}. By contrast, $\lambda_L$ does not fully saturate the bound for a critical Fermi surface~\cite{patel_quantum_2016} and is parametrically smaller in Fermi liquids or other weakly coupled states~\cite{stanford_many-body_2016, aleiner_microscopic_2016, banerjee_solvable_2016}.

Here, we study both time-ordered and OTO correlators in a diffusive many-body system by considering the concrete example of the non-integrable, one-dimensional Bose-Hubbard model. Thus far, it is a largely open question, how OTO correlators spread in diffusive systems with a few globally conserved quantities~\cite{hosur_chaos_2016, aleiner_microscopic_2016, patel_quantum_2016, mezei_entanglement_2016}. In our work, we study this question by performing matrix-product operator (MPO) based simulations of the Bose-Hubbard model at high temperatures,  at which well defined quasi-particles cease to exist. We demonstrate that in this regime the time-ordered one-particle correlation functions are strongly incoherent and feature rapidly decaying excitations, whereas the OTO correlators indeed describe the \emph{ballistic} spreading of information across the quantum system (see \fig{fig:otoc}). In contrast to the linear light-cone spreading of quantum information, the eventual \emph{global} thermalization of the closed system takes parametrically longer, due to hydrodynamic power-laws resulting from globally conserved quantities. For example,  we show that the local density correlation function decays as $\sim 1/\sqrt{Dt}$, describing diffusion in one dimension with the corresponding diffusion constant $D$. Thus, the time scales associated with the spread of information and with global thermalization are different.

Despite their usefulness to characterize interacting many-body systems theoretically, it remains a challenge to experimentally measure such dynamical correlation functions in \emph{real} space and time~\cite{knap_probing_2013,serbyn_interferometric_2014}, as required to observe information spreading. 
Here, we propose generic experimental protocols to characterize both time-ordered and OTO correlators via \textit{local} many-body interferometry. Our proposal to measure OTO correlators is unique because it overcomes some of the challenges that recently proposed protocols exhibit at finite temperatures and because it eliminates the scaling problems associated with \textit{global} many-body interferometry~\cite{islam_measuring_2015, kaufman_quantum_2016}. Furthermore, our protocol does not require an ancillary atom to switch between different system Hamiltonians~\cite{swingle_measuring_2016, zhu_measurement_2016, yao_interferometric_2016} and directly works with massive bosonic and fermionic particles (see also~\cite{shen_out--time-order_2016, tsuji_exact_2016, roberts_lieb-robinson_2016, halpern_jarzynski-like_2016, campisi_thermodynamics_2016}). We show that our protocol not only enables the measurement of dynamical correlation functions but also rather generic static correlation functions (including off-diagonal ones), thus opening the way for a full state-tomography of many-body quantum states. 

\begin{figure}
  \centering 
  \includegraphics[width=.48\textwidth]{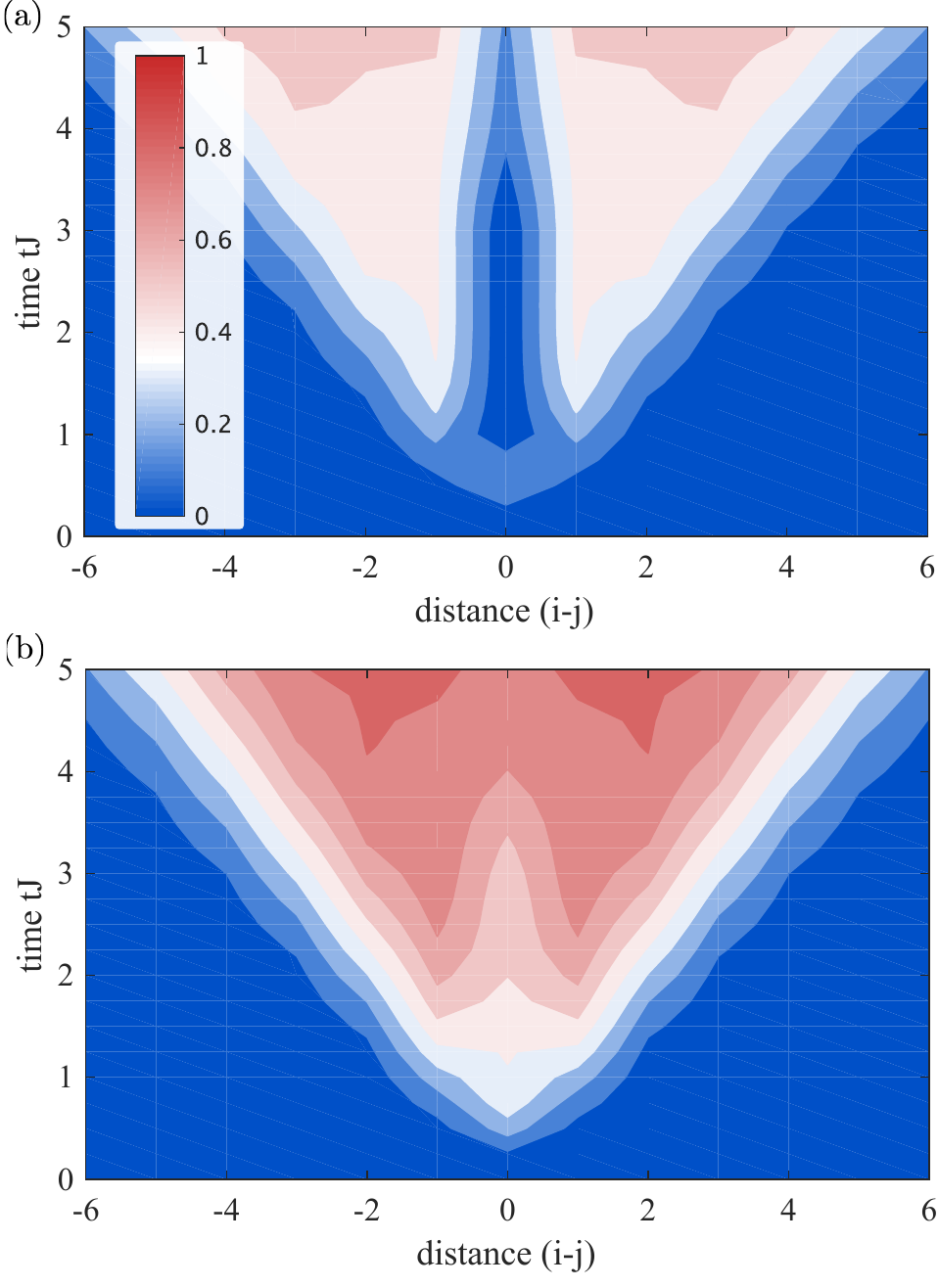}
  \caption{ \textbf{Light-cone spreading of quantum information.} Contour plots of the reduced OTO correlator $\mathcal{F}^r_{ij}(t)\sim |\mathcal{F}_{ij}(t) - \langle \hat n_i \hat n_j \rangle|/\langle \hat n_i \hat n_j \rangle$ as a function of time and distance for interaction strength $U=J$, chemical potential $\mu=0$, and temperature \fc{a} $T=2J$ and \fc{b} $T=16J$, respectively. The spreading of quantum information forms a light-cone pattern. The contour lines indicate changes of $\mathcal{F}^r_{ij}(t)$ by 0.1. 
  } 
  \label{fig:otoc_contour}
\end{figure}
\begin{figure*}
  \centering 
  \includegraphics[width=.9\textwidth]{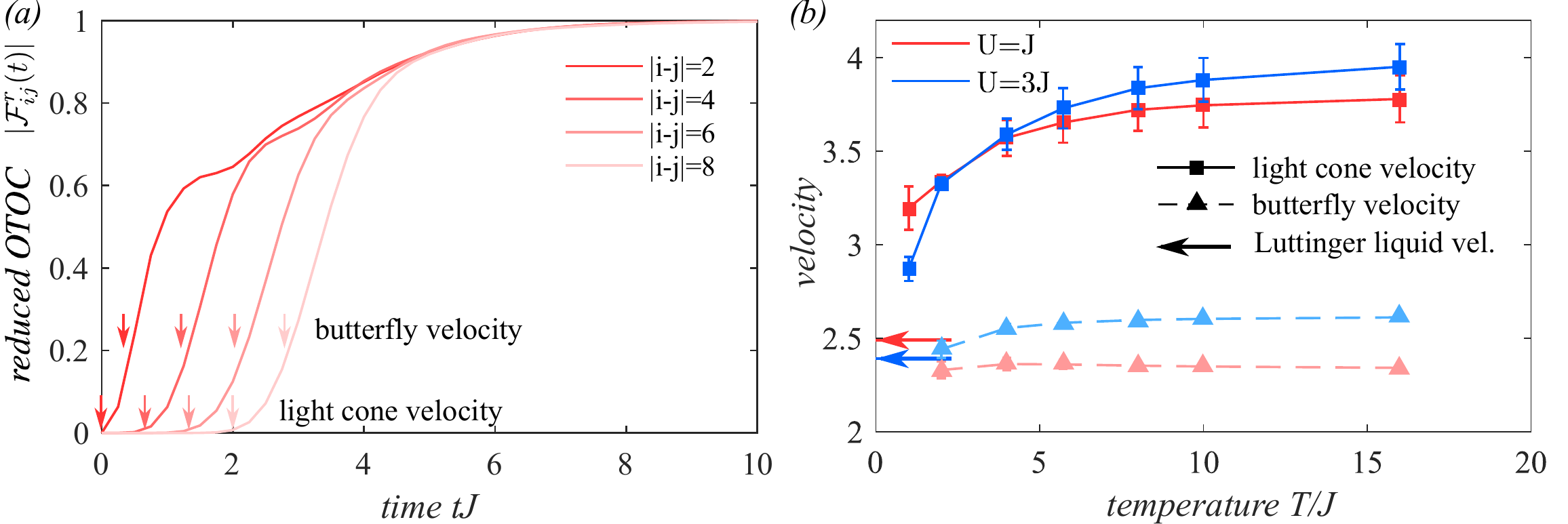}
  \caption{ \textbf{Characterizing the speed of information propagation.} \fc{a} Reduced OTO correlators $\mathcal{F}^r_{ij}(t)$ are shown as a function of time for different distances $|i-j|$, interaction strength $U=J$, and temperature $T=4J$. We introduce the light-cone velocity $v_\text{lc}$ by the space-time region, where $\mathcal{F}^r$ surpasses a small threshold and the butterfly velocity $v_\text{b}$ where it attains a large fraction of order one. \fc{b} The light-cone velocity $v_\text{lc}$ grows with temperature and is bounded from below by the zero temperature Luttinger liquid velocity (colored arrows). By contrast, the butterfly velocity $v_\text{b}$ is systematically smaller than $v_\text{lc}$ and approximately independent of temperature $T$. The data is shown for two values of the interaction strength $U=J$ and $U=3J$. 
  } 
  \label{fig:otoc_velocity}
\end{figure*}

\section{Results}

We study dynamical correlation functions of the one-dimensional Bose-Hubbard model focusing mainly on the incoherent intermediate to high temperature regime. The Hamiltonian of the system is given by
\begin{align}
 \hat H =& -J \sum_i ( c_i^\dag c_{i+1}^\nag + \text{h.c.}) + \frac{U}{2} \sum_i \hat n_i (\hat n_{i}-1) - \mu \sum_i \hat n_i,
 \label{eq:h}
\end{align}
where $J$ is the tunneling matrix element, $U$ the interaction strength, and $\mu$ the chemical potential. The bosonic creation (annihilation) operator on site $i$ is denoted as $c_i^\dag$ ($c_i$) and the local particle number operator is $\hat n_i=c_i^\dag c_i^\nag$. 

At zero temperature and commensurate filling, the Bose-Hubbard model exhibits a quantum phase transition from a gapped Mott insulating phase with short range correlations at strong interactions to a compressible superfluid phase with power-law correlations at weak interactions~\cite{giamarchi_quantum_2004}. At finite temperatures the system is a correlated, normal fluid. We compute the dynamical correlation functions at finite-temperature for systems up to $L=50$ sites using MPO techniques. The presented results are evaluated for virtual bond dimension $200$ to $400$ and the local bosonic Hilbert space is truncated to three states, which is sufficient to render the system nonintegrable. The presented results are checked for convergence with respect to the MPO bond dimension and system size; see Methods~\ref{app:convergence} for details on the numerical simulations.

\subsection{Spread of Quantum Information} 
\label{sec:spread}

Recently, OTO correlation functions have been proposed as a useful diagnostic tool to quantify the dynamical spreading of quantum entanglement and quantum chaos in many-body systems. OTO correlators describe the growth of the commutator between two local operators $\hat W$ and $\hat V$ in time
\begin{equation}
 C(t) = - \langle |[\hat W(t),\hat V(0)]|^2 \rangle.
 \label{eq:otoc1}
\end{equation}
In a semiclassical picture, the commutator in \eq{eq:otoc1} can be replaced by Poisson brackets. Then, for the choice of $\hat W=p_j$ and $\hat V=p_i$, this quantity reduces to $C(t) \sim \langle (\partial p_j(t) / \partial q_i(0))^2 \rangle$. Therefore, the correlation function $C(t)$ describes the sensitivity of the time evolution and is expected to grow exponentially at short times $\sim \exp[ \lambda_L t]$, with a rate $\lambda_L$ that resembles the Lyapunov exponent in classical chaotic dynamics. Rewriting these momenta and coordinates as combinations of creation and annihilation operators, \eq{eq:otoc1} generically consists of OTO correlators of the form
\begin{equation}
 \mathcal{F}_{ij}(t)=\langle c_j^\dag(t) c_i^\dag c_j(t) c_i \rangle.
 \label{eq:otoc}
\end{equation}
Below we mainly consider the quantum statistical average $\langle \ldots \rangle = \tr[\hat \rho \ldots]$ over an initial thermal state with weights distributed according to the Gibbs ensemble $\hat\rho = e^{-\hat H/T}/Z$ where $Z$ is the partition function and we set the Boltzmann constant $k_B$ to one. Alternatively, the average can also be performed with respect to an arbitrary initial state, for example a pure state $\hat\rho=\ket{\psi_0}\bra{\psi_0}$. For thermalizing systems, it is then expected that an effective temperature is approached at late times which depends on the energy density imprinted on the system by the initial state~\cite{deutsch_quantum_1991,srednicki_chaos_1994,rigol_thermalization_2008}.

OTO correlators $\mathcal{F}_{ij}$ evaluated at comparatively high temperatures $T=4J$, interactions $U=J$, and chemical potential $\mu=0$ are shown in \figc{fig:otoc}{a} as a function of time $t$ and distance $(i-j)$. Despite the high temperature, the OTO correlator $\mathcal{F}_{ij}$ unveils a pronounced light-cone spreading of the information across the quantum state for $|i-j|\lesssim 7$. For larger distances the light cone seems to grow super-ballistically, which we, however, attribute to the finite MPO bond dimension considered in the numerical simulations; see Methods~\ref{app:convergence}. OTO correlators are in that respect challenging to simulate with MPO techniques, because they directly reflect the fast spreading of entanglement.

The OTO correlator $\mathcal{F}_{ij}(t)$ should be contrasted to the time-ordered single-particle Green's function 
\begin{equation}
 \mathcal{G}_{ij}(t)=\langle c_j^\dag(t) c_i  \rangle,
 \label{eq:gf}
\end{equation}
which is shown in \figc{fig:otoc}{b}. In the incoherent transport regime, where well-defined quasiparticles do not exist, the Green's function $\mathcal{G}_{ij}(t)$ rapidly decays in time. Therefore, it is not capable of characterizing the spread of quantum information or entanglement across the state which is generically not linked  to the transport of quasi-particles~\cite{kim_ballistic_2013}. For the chosen parameters ($U=J$, $\mu=0$, $T=4J$), we find that the quasiparticle lifetime is approximately $\tau J \sim 0.6$ and hence shorter than the microscopic hopping rate, which indicates incoherent transport. By contrast, the out-of-time ordered structure of $\mathcal{F}_{ij}(t)$ reveals a well defined linear spread of quantum information despite the high temperature.

We now characterize the OTO correlators $\mathcal{F}_{ij}(t)$ in detail. To this end, we subtract $\langle \hat n_i \hat n_j \rangle$ from the $\mathcal{F}_{ij}(t)$ and consider its relative change: $\mathcal{F}_{ij}^r(t)=|\mathcal{F}_{ij}(t)-\langle \hat n_i \hat n_j \rangle|/\langle \hat n_i \hat n_j \rangle$. Examples for the reduced OTO correlator $\mathcal{F}_{ij}^r(t)$ are shown in \fig{fig:otoc_contour} for interaction $U=J$ and different temperatures $T$. The reduced OTO correlator $\mathcal{F}_{ij}^r(t)$ starts off at zero, forms the light-cone plateau, and approaches the steady-state value as an exponential. 

From the light-cone spread of the OTO correlator, we extract two velocities [\figc{fig:otoc_velocity}{a}]: (1) The light-cone velocity $v_\text{lc}$, which we define by the space-time region where the reduced OTO correlators $\mathcal{F}_{ij}^r(t)$ surpasses a small threshold of 0.05\% of its final value. (2) The butterfly velocity $v_\text{b}$, which we define by the space-time region where the OTO correlator attains a large fraction (20\%) of its final value. We find that $v_\text{b}$ does not significantly depend on this cutoff, as long as it is chosen to a sizeable fraction; see Methods \fig{fig:velocities}. The light cone velocity $v_\text{lc}$ increases with temperature $T$ and is bounded from below by the zero temperature Luttinger liquid velocity; see \figc{fig:otoc_velocity}{b}. The butterfly velocity $v_\text{b}$ is systematically lower than $v_\text{lc}$ and is almost independent of temperature. The butterfly velocity determines the time scale $t_\text{scr}$ for scrambling information across the many-body quantum state which is linear in system size $t_\text{scr}\sim L/v_\text{b}$. Based on results from holography, it has been argued in Ref.~\cite{mezei_entanglement_2016} that the light-cone and the butterfly velocity should be quite generally the same. This should be contrasted to our results for the Bose-Hubbard model and to a study of non-relativistic non-Fermi liquids~\cite{patel_quantum_2016}. In both cases the butterfly velocity has been found to be smaller than the light-cone velocity. For details on the analysis of the OTO correlators; see Methods~\ref{app:data}. 

\begin{figure}
  \centering 
  \includegraphics[width=.45\textwidth]{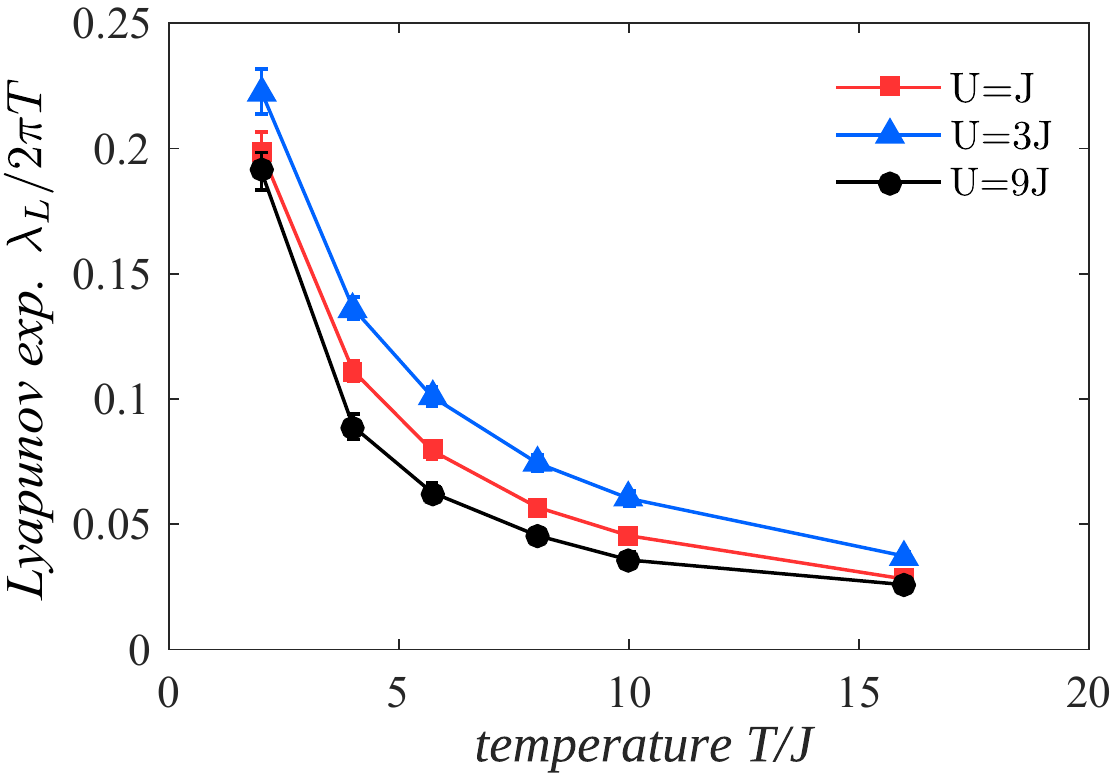}
  \caption{ \textbf{Lyapunov exponent.} The reduced OTO correlator $\mathcal{F}^r_{ij}(t)$ is expected to grow exponentially on a timescale set by the butterfly velocity $v_\text{b}$ with a rate that defines the Lyapunov exponent $\lambda_L$. In our system, the regime of exponential growth is restricted to a rather small time range, see also \fig{fig:lyapunov}. Our data suggests that the Lyapunov exponent $\lambda_L$ is parametrically smaller than the conjectured upper bound $2\pi T$ and increases slowly as the temperature $T$ is lowered. The data is shown for interaction strength $U=\{1, 3, 9\}J$.
  } 
  \label{fig:otoc_Lyapunov}
\end{figure}

We characterize the initial dynamics of the OTO correlator by its growth rate $\lambda_L$ around the space-time cone set by the butterfly velocity: $\mathcal{F}^r_x(t)\sim \exp[\lambda_L (t-x/v_\text{b})]$ and refer to this rate as Lyapunov exponent. 
However, we note that in contrast with predictions from strongly coupled field theories~\cite{maldacena_bound_2016} or disordered SYK models~\cite{kitaev_2014}, there appears to be no parametrically large regime of exponential growth in our model, see \fig{fig:lyapunov}, which hints at more complex initial dynamics. Finding the analytic form for the initial growth over an extended space-time region remains an outstanding challenge. Alternatively, one can extract the growth rate $\Lambda_\text{lc}$ of the reduced OTO correlator around a space-time region set by the light-cone velocity $v_\text{lc}$ which we find to be larger than $\lambda_L$; see Methods~\ref{app:data}. We show the Lyapunov exponent $\lambda_L$ as a function of temperature for different values of the interaction strength $U$ and chemical potential $\mu=0$  in \fig{fig:otoc_Lyapunov}. It has been conjectured that the Lyapunov exponent is bounded by $2\pi T$, which is the value it assumes in a strongly coupled field theory with a gravity dual~\cite{maldacena_bound_2016}. In our system, $\lambda_L$ is parametrically lower than this bound and increases slowly when lowering the temperature. Moreover, we find that the dependence of the Lyapunov exponent on the interaction strength $U$ is small with slightly larger values of $\lambda_L$ for intermediate interaction strength, $U=3J$, which is in the vicinity of the quantum critical point. 

\subsection{Thermalization}

\begin{figure}
  \centering 
  \includegraphics[width=.48\textwidth]{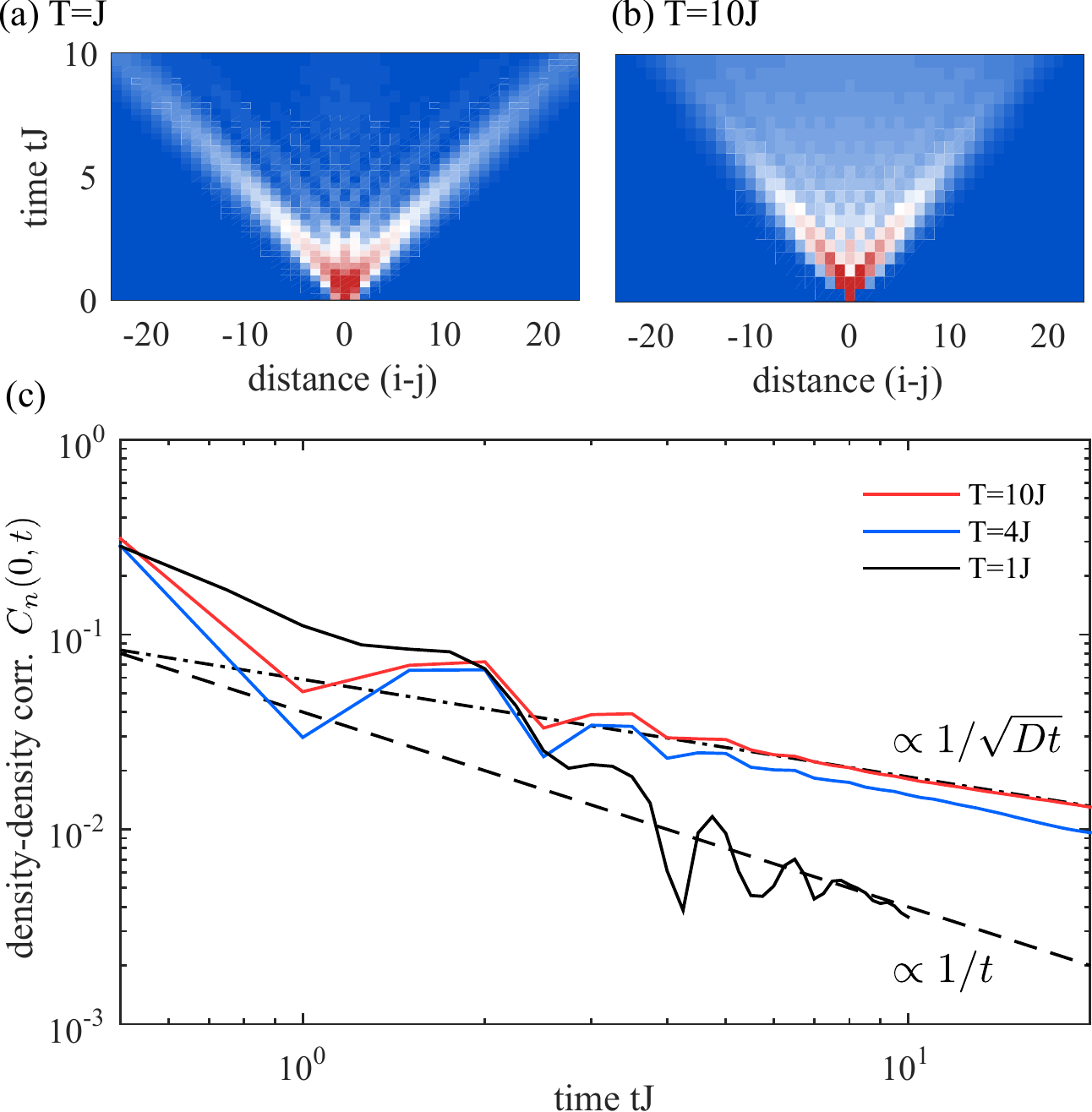}
  \caption{ \textbf{Thermalization in closed quantum systems.} Conserved quantities restrict the approach of a closed quantum system to global equilibrium, thus, rendering global thermalization a slow process. In the Bose-Hubbard model the total particle number is conserved leading to diffusive power-law tails in the connected density correlator $C_{n}(x,t) = \re [ \langle \hat n_x(t) \hat n_0 \rangle -\langle \hat n_x\rangle \langle \hat n_0\rangle ]$. \fc{a} At low temperatures ($T=J$), where quasiparticles are reasonably well defined, the density correlator does not reach the diffusive regime within the accessible simulation time but is dominated by ballistic sound peaks. \fc{b} By contrast, for high temperatures ($T=10J$) the crossover to diffusion becomes apparent. \fc{c} For temperatures $T\gtrsim 4J$ the local density correlator $C_{n}(0,t)\sim 1/\sqrt{Dt}$, where $D$ is the diffusion constant. By contrast, at low temperature $T=J$ the diffusive regime has not yet been reached within the numerically accessible times and the correlations rather decay ballistically $C_{n}(0,t)\sim 1/{t}$. The slow relaxation of the hydrodynamic modes leads to the global thermalization time scale $t_\text{th} \sim L^2/D$ that is parametrically larger than the scrambling time scale $t_\text{scr} \sim L/v_\text{b}$ of quantum information.
  } 
  \label{fig:therm}
\end{figure}

Closed quantum systems approach their global equilibrium only very slowly, due to the slow evolution of observables that overlap with conserved quantities. In the Bose-Hubbard model \eqw{eq:h}, energy, lattice momentum, and total particle number are conserved. From hydrodynamics we infer that, for example, the conserved particle number leads to a diffusion equation of the density~\cite{chaikin_principles_2000,lux_hydrodynamic_2014}
\begin{equation}
 \partial_t n - D \nabla^2 n = 0,
\end{equation}
where $D$ is the diffusion constant. The connected density correlation function $C_{n}(x-x',t-t')=\re [\langle n(x,t) n(x',t') \rangle - \langle n(x) \rangle \langle n(x') \rangle]$ relates the density at space-time $(x,t)$ to the density at $(x',t')$ via $n(x,t) \sim \int dt' dx' C_{n}(x-x',t-t') n(x',t')$ and in a hydrodynamic regime is expected to be of the form
\begin{equation}
 C_{n}(x,t) \cong \frac{\tilde C}{\sqrt{4 \pi D |t|}} e^{-\frac{x^2}{4D|t|}},
\end{equation}
with $\tilde C = \int dx \, C_{n}(x,0)$. Whereas local equilibrium is approached after a few scattering events, attaining global equilibrium is restricted due to the relaxation of such conserved quantities, which have to be transported over long distances. At comparatively low temperatures ($T=J$), \figc{fig:therm}{a}, the ballistic spread of sound modes dominates the dynamics of the connected density correlator in the numerically accessible time regime. However, at high temperatures ($T=10J$), \fc{b}, the density correlator approaches diffusive transport after a few hopping scales and attains a finite value in the region between the sound modes. To be more quantitative, we study the local ($x-x'=0$) density correlation function. At high temperatures $T\gtrsim 4J$ the local correlator exhibits a diffusive power-law decay $C_{n}(x=0,t)\sim 1/\sqrt{D t}$, \fc{c}. For this parameter set we extract the diffusion constant $D= 9.79(1) a^2J$ for $T=10J$ and $D= 14.29(27) a^2J$ for $T=4J$, where $a$ is the lattice spacing; see \tab{tab:diffusion_constants}. The decrease of the diffusion constant with increasing temperature is somewhat counterintuitive. We attribute this behavior to the fact that the calculations are performed in the grand-canonical ensemble. Hence the particle density depends on the temperature and, in particular, increases with temperature in the chosen parameter regime. 
We note that the connected density correlator does not exhibit pronounced hydrodynamic long time tails, which could result from higher order gradient corrections to the diffusion equation and mask the $1/\sqrt{Dt}$ decay. This seems to be a particular property of the density correlator, as we find at high-temperatures pronounced $t^{-3/4}$ corrections in the energy-density correlation function (not shown), in agreement with Ref.~\cite{lux_hydrodynamic_2014}.

\begin{table}
\centering
\begin{tabular}{rlrlr}
\hline
\multicolumn{1}{l}{$\quad\;\; T/J$} & $\qquad$ & \multicolumn{1}{l}{$D/(a^2 J)$} & $\qquad$ & \multicolumn{1}{l}{$D\lambda_L/v_b^2$} \\
\hline
$\qquad$4 $\quad$                      &          & 14.29(27) $\quad$                           &          & 7.2(6) $\quad$                             \\
$\qquad$6 $\quad$                      &          & 11.69(10) $\quad$                           &          & 6.0(4)  $\quad$                               \\
$\qquad$8  $\quad$                     &          & 10.42(04) $\quad$                           &          & 5.4(3)   $\quad$                              \\
$\qquad$10 $\quad$                     &          & 9.79(01)  $\quad$                           &          & 5.1(3)  $\quad$                               \\
\hline
\end{tabular}
\caption{\textbf{Diffusion constant $D$ and the ratio $D \lambda_L/v_B^2$ for different temperatures $T$.} The errors as indicated in the parentheses are errors from the fits.}
\label{tab:diffusion_constants}
\end{table}

It has been proposed that the diffusion constant is related to the butterfly velocity $v_\text{b}$ and the Lyapunov exponent $\lambda_L$ via $D \sim v_\text{b}^2/\lambda_L$~\cite{blake_universal_2016, hartnoll_theory_2015, swingle_slow_2016, lucas_charge_2016}, where $1/\lambda_L$ is a bound for the local thermalization time in which the system is able to attain local equilibrium characterized by a local temperature and local chemical potential that varies between different regions in space. From our simulations, we obtain coefficients of the order $D \lambda_L/v_\text{b}^2 \sim 5.5$ for temperatures $T\gtrsim 6J$; see Tab.~\ref{tab:diffusion_constants}, which seems to suggest a connection between the spread of information and \emph{local} thermalization, as suggested by calculations for holographic matter. However, clearly \emph{global} thermalization is a parametrically slower process than information scrambling and takes for systems of size $L$ times of the order $t_\text{th}\sim L^2/D$. Experimentally measuring OTO correlators (Sec.~\ref{sec:measurement}) and density correlators (App.~\ref{app:dens}) will make it possible to further check these holographic predictions.

\subsection{Measuring Dynamical Time-Ordered and Out-of-Time Ordered Correlators\label{sec:measurement}}

We  develop two generic interferometric protocols that measure time-ordered as well as OTO correlation functions for systems of bosons or fermions in an optical lattice. The first is based on \emph{globally} interfering two many-body states and the second on \emph{local} interference. The quantum interference of two copies of the many-body state is realized by local beam splitter operations. Variants of this approach have been proposed to study R{\'e}nyi entropies~\cite{moura_alves_multipartite_2004, daley_measuring_2012, pichler_thermal_2013} and have been demonstrated experimentally using a quantum gas microscope~\cite{islam_measuring_2015, kaufman_quantum_2016}. Both protocols that we propose consist only of elements which have already been used in experiments. The two protocols are complementary and each of them has its own advantages. 

Global interferometry precisely yields the square modulus of the single-particle Green's functions $\mathcal{G}_{ij}^\text{gl}(t) = |\mathcal{G}_{ij}(t)|^2$ and the OTO correlators $\mathcal{F}^\text{gl}_{ij}(t) = |\mathcal{F}_{ij}(t)|^2$ for \emph{pure} initial states (see Supplemental Information Sec.~\ref{sec:global}). In a thermalizing system, effective finite temperatures can be obtained with the help of quenches from pure initial states. However, generic high temperature initial states are not accessible with this protocol. The measurement schemes proposed e.g. in Ref.~\cite{yao_interferometric_2016} face similar challenges.
The global interferometry protocol is furthermore limited to rather small system sizes, since the many-body wave function overlap has to be measured, which requires an extensive number of beam splitter operations. 

These limitations are overcome by the second proposed protocol which uses local interferometry (see Supplemental Information Sec.~\ref{sec:local} for details on the implementation). In this protocol, only two local beam-splitter operations are required irrespective of the system size, and only local density differences between the two copies have to be measured. Furthermore, initial thermal density matrices can be studied as well.  This local approach yields a slightly amended two-point correlation function $\mathcal{G}^\text{loc}_{ij}(t) \sim \im [ \langle a_j^\dag(t) a_i \rangle \langle a_j(t) a_i^\dag \rangle ]$ and OTO correlator $\mathcal{F}^\text{loc}_{ij}(t) \sim \im [ \langle a_j^\dag(t) a_i^\dag a_j(t) a_i \rangle \langle a_j^\dag(t) a_i a_j(t) a_i^\dag \rangle ]$. However, we demonstrate in the Supplemental Material that these correlators carry much of the same information as the ones we discussed previously. Static correlation functions are accessible with the same techniques and an extension to higher order correlators is straight forward.

\section{Discussion}

We studied time-ordered as well as out-of-time ordered correlation functions in the one-dimensional Bose-Hubbard model and suggest different protocols to experimentally access them. At high temperatures, well-defined quasi-particles cease to exist and the time-ordered Green's function decays within short times. However, the spread of information is not necessarily linked to the transport of quasi-particles. Our numerical results for the out-of-time ordered correlators clearly indicate the ballistic spread of information even at high temperatures where transport is incoherent. In our one-dimensional system, this linear spread sets the timescale for scrambling information across the quantum state to be proportional to the system size. Moreover, the existence of conserved quantities in the Bose-Hubbard model leads to diffusive behavior of the corresponding time-ordered correlation functions. Global thermalization therefore scales with the square of the system size and takes parametrically longer than scrambling quantum information.

For future work, it would be on the one hand interesting to develop analytical predictions for the growth of OTO correlators, which in our numerics deviates significantly from the simple exponential growth obtained in strongly coupled field theories, or for the bounds that characterize the information propagation and Lyapunov exponents. On the other hand, the numerical study of out-of-time ordered correlators in other interacting many-body systems, including Fermi-Hubbard models, spin models, or continuum Lieb-Liniger models, could be beneficial. Taking such routes could help to advance our fundamental understanding of information scrambling, transport, and thermalization.

Experimental measurements of both time-ordered and out-of-time ordered correlators will be eminent for the investigation of the dynamical properties of many-body systems. We proposed two different protocols to measure correlation functions, which can be either static, time ordered, or out-of-time ordered. The schemes are respectively based on the global and local interference of two copies of the many-body state of interest. The required techniques have already been demonstrated in experiments with synthetic quantum matter. An extension of the described experimental schemes to two-dimensional systems is conceivable as well and could provide a valuable perspective on the dynamics of many-body quantum systems.

\section{Methods}

In this section, we discuss the numerical method with which the calculations have been performed as well as the procedure to obtain the velocities $v_{lc}$, $v_b$ and the Lyapunov exponent $\lambda_L$. 

\subsection{Numerical Simulations}
\label{app:convergence}

\begin{figure*}
  \centering 
  \includegraphics[width=.98\textwidth]{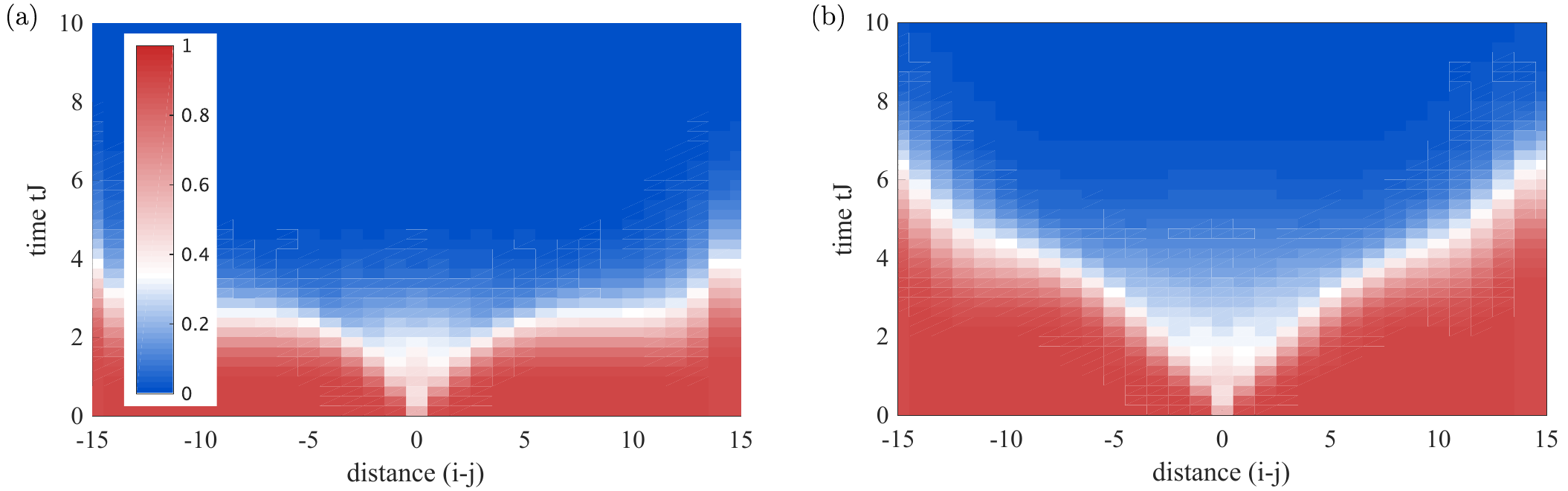}
  \caption{ \textbf{Comparison of numerical data for different bond dimensions.} OTO correlators $\mathcal{F}_{ij}(t)$ are shown for $T=4J$ and $U=J$, and bond dimension \fc{a} $20$ and \fc{b} $400$. The plateau emerging around $|i-j| \approx 5$ in \fc{a} diminishes and appears only at larger distances for the higher bond dimension shown in \fc{b}. However, despite the large difference in the bond dimension, deviations from the linear light cone are still apparent.
  } 
  \label{fig:comp_maxD}
\end{figure*}

Our numerical simulations are based on finite-temperature, time-dependent matrix product operators (MPO)~\cite{tDMRG2004, VerstraetePRL2004, VerstraeteMurgCirac2008, Schollwock2011, Barthel2013, Karrasch2013}. 

For the density correlations, we evaluate~\cite{Barthel2013}
\begin{equation}
\begin{split}
\label{eq:nn_regroup}
&\big\langle \hat{n}_{\ell}(t) \hat{n}_j(0) \big\rangle_{\beta} \stackrel{\text{def}}{=} \frac{1}{Z} \tr\!\left[ \mathrm{e}^{-\beta \hat{H}} \big(\mathrm{e}^{i t \hat{H}} \hat{n}_{\ell} \mathrm{e}^{-i t \hat{H}}\big) \hat{n}_j\right] \\
& = \frac{1}{Z} \tr\!\left[ \big( \mathrm{e}^{i \frac{t}{2} \hat{H}} \mathrm{e}^{-\beta \hat{H}/2} \hat{n}_{\ell} \mathrm{e}^{-i \frac{t}{2} \hat{H}}\big) \big(\mathrm{e}^{-i \frac{t}{2} \hat{H}} \hat{n}_j \mathrm{e}^{-\beta \hat{H}/2} \mathrm{e}^{i \frac{t}{2}\hat{H}}\big) \right],
\end{split}
\end{equation}
where $\beta$ is the inverse temperature and $Z$ the partition function. We construct the MPO approximation of the two terms in the parentheses by first computing $\mathrm{e}^{-\beta \hat{H}/2} \hat{n}_{\ell}$ and $\hat{n}_j \mathrm{e}^{-\beta \hat{H}/2}$, respectively, and then performing a real-time evolution up to $\frac{t}{2}$ and $-\frac{t}{2}$. By exploiting the time translation invariance, $\langle \hat{n}_{\ell}(t) \hat{n}_j(0)\rangle_{\beta} = \langle \hat{n}_{\ell}(t/2)\,\hat{n}_j(-t/2)\rangle_{\beta}$, the maximum simulated time has effectively been reduced by a factor two, which in turn reduces the required virtual bond dimension of the MPO.

To evaluate $\mathrm{e}^{-\beta \hat{H}/2}$, we employ a second-order Suzuki-Trotter decomposition with imaginary time step $\Delta\tau$ (typically $\Delta\tau J = 0.025$) after splitting the Hamiltonian into even and odd bonds, as described in Ref.~\onlinecite{tDMRG2004}. The real-time evolution proceeds by Liouville steps $\hat{A}(t+\Delta t) = \mathrm{e}^{i \Delta t \hat{H}} \hat{A}(t) \mathrm{e}^{-i \Delta t \hat{H}}$. For each of the steps we combine a fourth-order partitioned Runge-Kutta method \cite{BlanesMoan2002} with even-odd bond splitting of the Hamiltonian. As noted in Ref.~\onlinecite{Barthel2013}, the Liouville time evolution has the advantage that the virtual bond dimension does not increase outside the space-time cone set by Lieb-Robinson-type bounds. The high order decomposition also allows for relatively large time steps (in our case $\Delta t J = 0.125$ or $0.25$).

For the OTO correlators $\langle c_j^\dag(t) c_{\ell}^\dag c_j(t) c_{\ell} \rangle_{\beta}$, a regrouping analogous to Eq.~\eqref{eq:nn_regroup} would lead to four terms inside the trace, such that a straightforward contraction to evaluate the trace becomes computationally very expensive. Instead, we evaluate
\begin{equation}
\begin{split}
&\big\langle c_j^\dag(t) c_{\ell}^\dag c_j(t) c_{\ell} \big\rangle_{\beta} \\
& = \frac{1}{Z} \tr\!\left[ \big(\mathrm{e}^{i t \hat{H}} \mathrm{e}^{-\beta \hat{H}} c^\dag_j \mathrm{e}^{-i t \hat{H}} c^\dag_{\ell}\big) \big(\mathrm{e}^{i t \hat{H}} c_j \mathrm{e}^{-i t \hat{H}} c_{\ell}\big) \right]
\end{split}
\end{equation}
and time-evolve both $\mathrm{e}^{-\beta \hat{H}} c^\dag_j$ and $c_j$ up to time $t$. Subsequent application of the site-local operators $c^\dag_{\ell}$ and $c_{\ell}$ does not affect the virtual bond dimension in the MPO representation.

In our simulations, we restrict the local Hilbert space to three states due to computational limitations. Since the average particle number per site is approximately one, this restriction should not qualitatively affect the simulation results. Moreover, truncating the local Hilbert space to three states is sufficient to render the system non-integrable, which is crucial to observe the thermalization behavior studied in this work.

Since OTO correlators are closely linked to the spreading of entanglement, it is challenging to simulate them using MPO techniques.  In \fig{fig:comp_maxD} we compare the data obtained for the same simulation parameters but different maximal bond dimensions. The MPO bond dimension of 20 leads to an apparently super-ballistic growth of the light-cone around $|i-j| \approx 5$, see \figc{fig:comp_maxD}{a}. Increasing the bond dimension shifts this numerical artifact to larger distances. It is however exponentially costly to reach full convergence of the OTO correlator. In the analysis of the numerical data we therefore only considered small distances, where we checked that increasing the bond dimension does not alter the correlators. 

\subsection{Data Analysis \label{app:data}}

\begin{figure}
  \centering 
  \includegraphics[width=.48\textwidth]{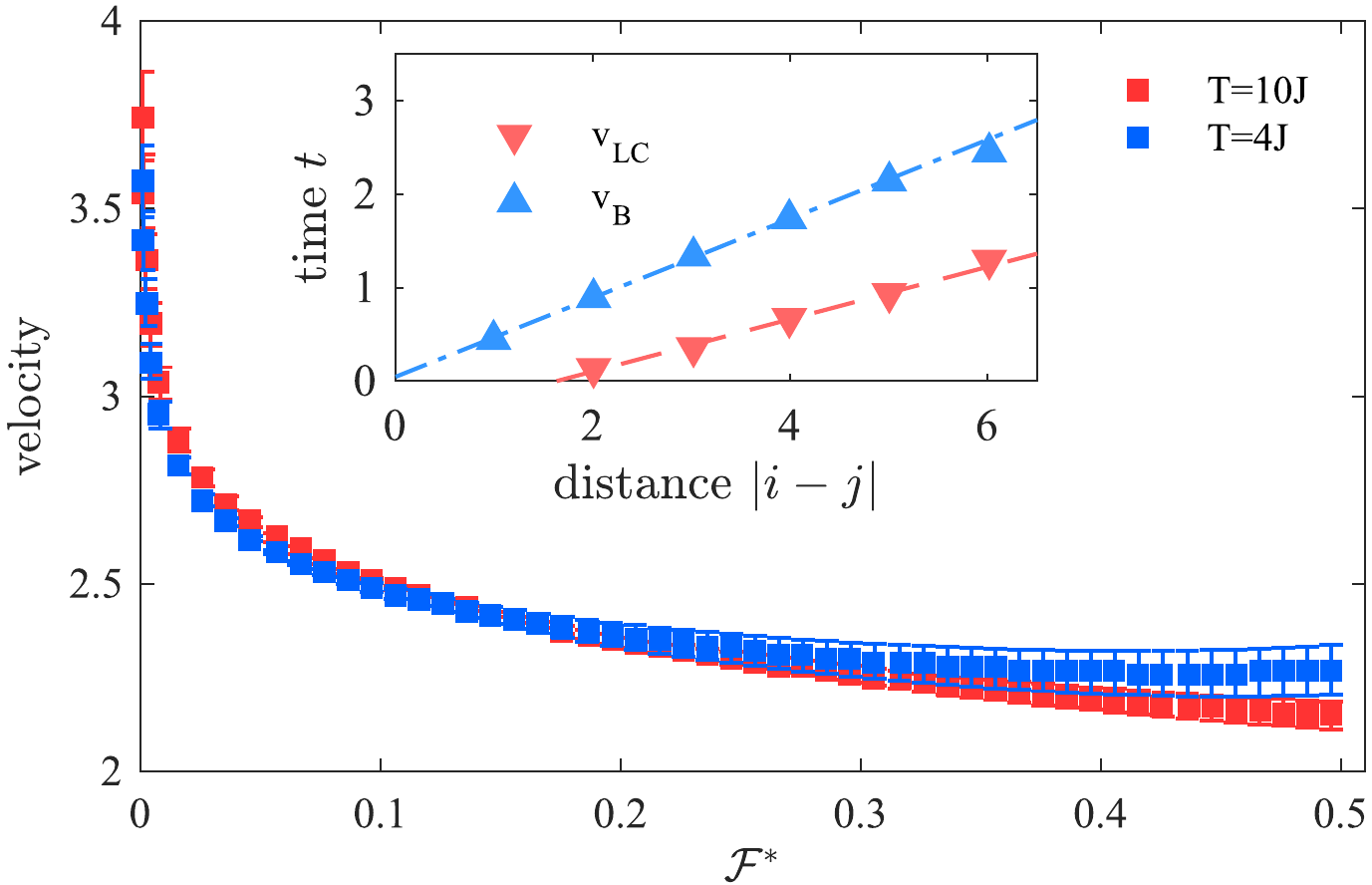}
  \caption{ \textbf{Determining the light-cone  $v_\text{lc}$ and butterfly velocity $v_\text{b}$.} The velocities resulting from a linear fit to the times at which $\mathcal{F}_{ij}^r(t)$ reaches the value $\mathcal{F}^*$ are shown for $U=J$ and $T=\{4,10\}J$. The errorbars are the fitting errors. Inset: The inverse slope of the linear fit to the times as a function of distance $|i-j|$ determines the different velocities (shown for $U=J$ and $T=4J$).
  } 
  \label{fig:velocities}
\end{figure}

\begin{figure*}
  \centering 
  \includegraphics[width=.98\textwidth]{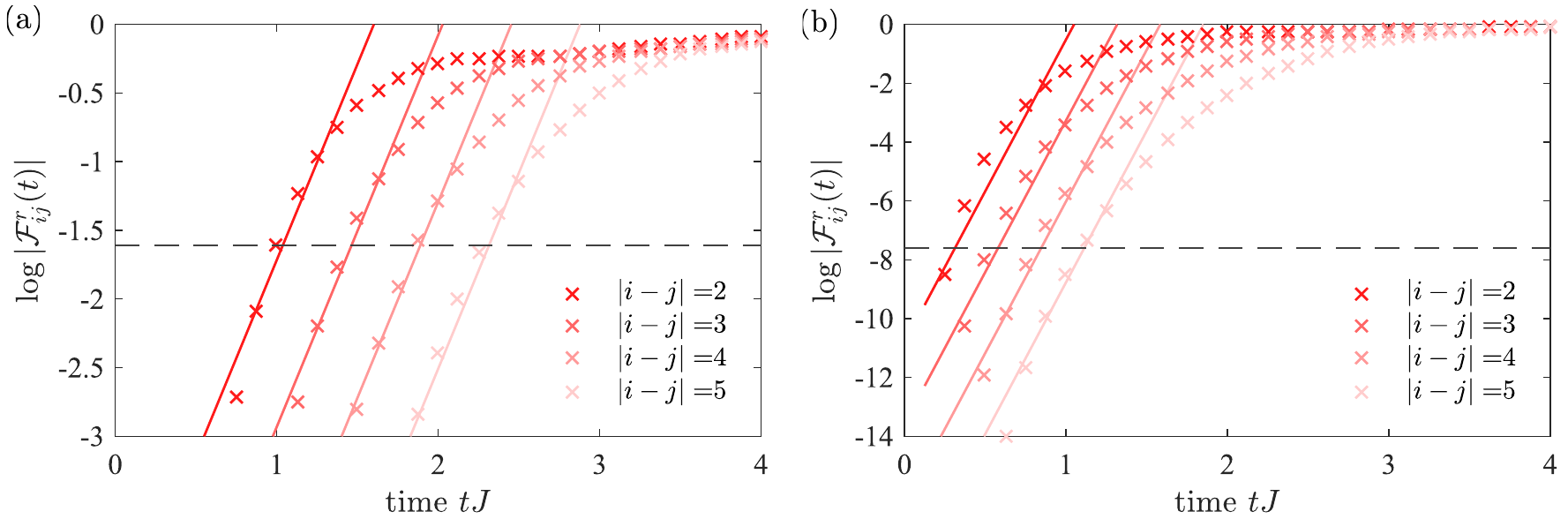}
  \caption{ \textbf{Determining the Lyapunov exponent $\lambda_L$.} OTO correlators $\mathcal{F}_{ij}^r(t)$ are shown for $T=10J$ and $U=J$. Solid lines depict the predicted exponential growth from which we determine \fc{a} the Lyapunov exponent $\lambda_L=2.9(1)$ and \fc{b} the light-cone exponent $\Lambda_{lc}=10.3(5)$. The dashed gray line denotes the threshold value $\mathcal{F}^*$ used to determine the velocities $v_\text{b}$ and $v_\text{lc}$, respectively. We obtain the exponents by fitting our data in a restricted regime around the threshold value $\mathcal{F}^*$ to the predicted exponential growth, see text for details. However, we note that in our data the exponential growth is limited to a rather small time range. The errorbars shown in \fig{fig:otoc_Lyapunov} correspond to errors obtained from such fits.
  } 
  \label{fig:lyapunov}
\end{figure*}

We describe in detail, how we determine the light-cone velocity $v_\text{lc}$, the butterfly velocity $v_\text{b}$, and the Lyapunov exponent $\lambda_L$. The light-cone velocity is defined as the ratio of the distance $|i-j|$ and the time at which the reduced OTO correlator $\mathcal{F}_{ij}^r(t)$ reaches a small threshold. The butterfly velocity, however, sets a scale for the time it takes to scramble information over the system and is therefore defined via the time at which $\mathcal{F}_{ij}^r(t)$ attains a large value of order one. The specific threshold one chooses to determine the butterfly velocity is thus somewhat arbitrary. We illustrate the dependence of the velocity $v$ on the chosen threshold $\mathcal{F}^*$ of the reduced OTO correlator $\mathcal{F}_{ij}^r(t)$ in \fig{fig:velocities}. For large values of $\mathcal{F}^*$, the velocity converges toward a constant. Hence, the butterfly velocity will be largely insensitive to the precise choice of $\mathcal{F}^*$ as long as it is large enough. For the definition of $v_\text{b}$, we consider the specific value of $\mathcal{F}^*_\text{b}=0.2$.

In the limit $\mathcal{F}^* \rightarrow 0$, there is a strong dependence of $v$ on the choice of the threshold. The light-cone velocity $v_\text{lc}$ is defined by the fastest spread of information through the system and is determined by the reduced OTO correlator attaining a small value. To fulfill this definition, we fix $\mathcal{F}^*_\text{lc} = 0.0005$; see inset in \fig{fig:velocities}.

As described in Sec.~\ref{sec:spread}, the OTO correlator is expected to grow exponentially on a timescale set approximately by the butterfly velocity. We thus fit the exponential function
\begin{equation}
\mathcal{F}_{x}^r(t) = a \cdot e^{\lambda_L\left(t-\frac{x}{v_\text{b}} \right)}
\end{equation}
to the numerical data simultaneously for distances $1 \leq |i-j| \leq 5$ within the range $-2.5 \leq \log \mathcal{F}_{ij}^r(t) \leq -1$. The butterfly velocity $v_\text{b}$ is determined as described above with the threshold $\mathcal{F}^* = 0.2$, which lies well within the interval of the considered data points; see \figc{fig:lyapunov}{a} for an exemplary plot. 

We note that the exponential growth of the reduced OTO correlator is in our model limited to a rather small dynamical range. Extracting by contrast the growth rate $\Lambda_\text{lc}$ of the OTO correlator from the timescale set by the light-cone velocity within the range $-14 \leq \log \mathcal{F}_{ij}^r(t) \leq -4.5$, \figc{fig:lyapunov}{b}, yields larger rates. In particular, for the parameters shown in \fig{fig:lyapunov}, we obtain  $\lambda_L=2.9(1)$ and $\Lambda_\text{lc}=10.3(5)$, respectively.

\begin{acknowledgments}
We thank T. Barthel, X. Chen, S. Gopalakrishnan,  H.-C. Jiang, A. Kitaev, and S. Sachdev for useful discussions. We acknowledge support from the Technical University of Munich - Institute for Advanced Study, funded by the German Excellence Initiative and the European Union FP7 under grant agreement 291763  (A.B, M.K), the DFG grant No. KN 1254/1-1 (A.B, M.K), the Studienstiftung des deutschen Volkes (A.B.), and the Alexander von Humboldt foundation via a Feodor Lynen fellowship (C.M.).
\end{acknowledgments}

\input{otoc_v15.bbl}

\newpage

\begin{center}
{\Large Supplemental Information} 
\end{center}

\appendix

\section{Details on the Experimental Protocols}

We elaborate on the different protocols outlined in Sec.~\ref{sec:measurement}, which can be used to experimentally access the theoretical findings presented in this work.

\subsection{Global Many-Body Interferometry \label{sec:global}}

\begin{figure*}
  \centering 
  \includegraphics[width=.7\textwidth]{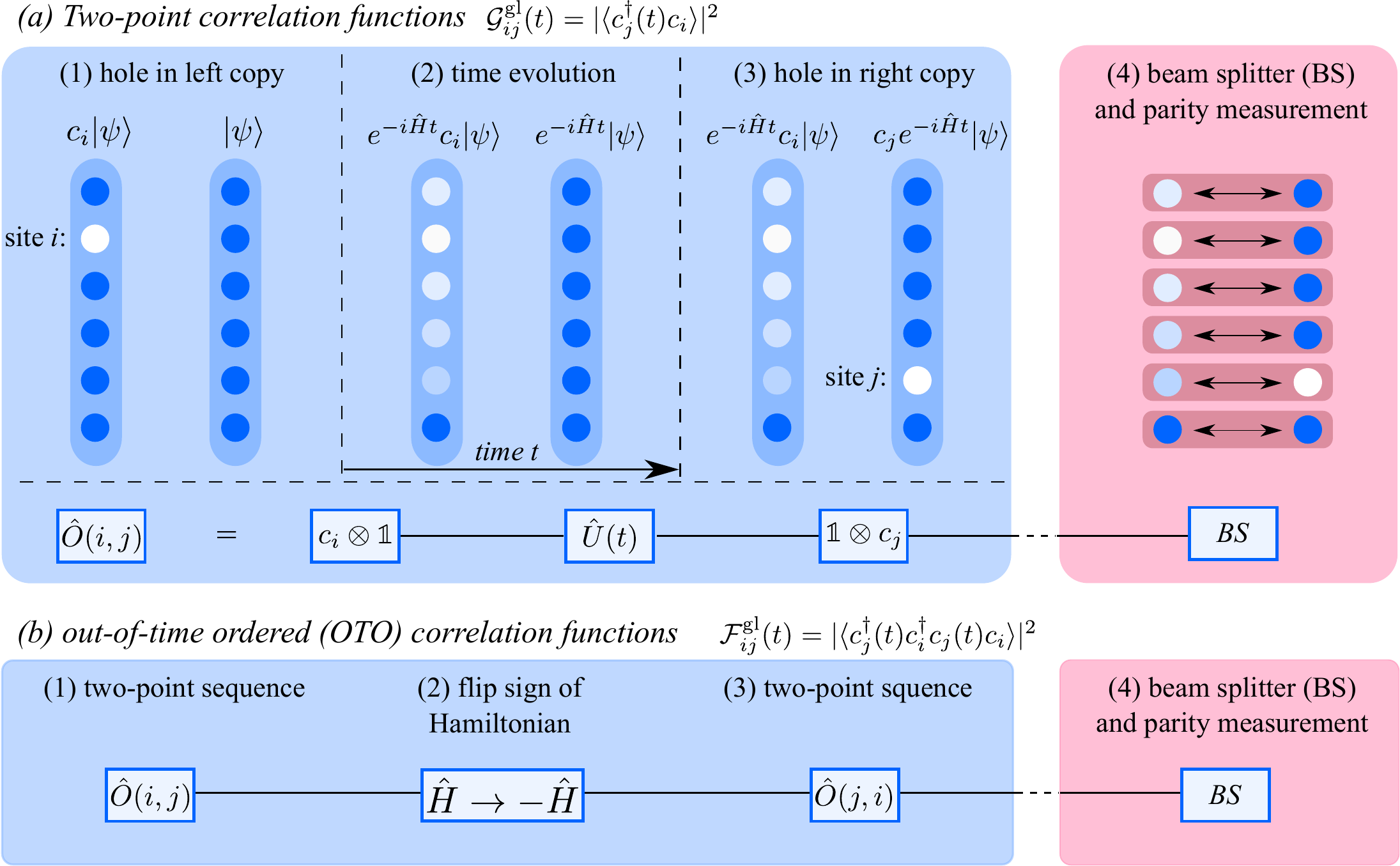}
  \caption{ \textbf{Probing dynamical correlation functions through the \emph{global} interference of two many-body states.} Schematic illustration of the experimental protocol to determine the \fc{a} time-ordered Green's function $\mathcal{G}^\text{gl}_{ij}(t)=|\bra{\psi}  c_j^\dag(t) c_i \ket{\psi}|^2$ as well as \fc{b} OTO correlation functions $\mathcal{F}^\text{gl}_{ij}(t)=|\bra{\psi}  c_j^\dag(t) c_i^\dag c_j(t) c_i \ket{\psi}|^2$. Details of the protocol are described in the text.   } 
  \label{fig:protocol}
\end{figure*}

We consider a system of bosons or fermions in an optical lattice. At this point, we do not make any assumptions about the specific form of the Hamiltonian $\hat H$. We first focus on the real-time and spatially resolved single-particle Green's functions $\mathcal{G}^\text{gl}_{ij}(t)$, which can be measured by the following protocol, \figc{fig:protocol}{a}: (1) Initially, prepare two identical copies of a pure state $\ket{\psi} \otimes \ket{\psi}$. Remove a particle on site $i$ in the left system by locally transferring the atom to a hyperfine state that is decoupled from the rest of the system or by transferring it to a higher band of the optical lattice, yielding $c_i^\nag \ket{\psi} \otimes \ket{\psi}$. (2) The system evolves in time for a period $t$, $\exp[-i \hat H t]c_i^\nag \ket{\psi} \otimes \exp[-i \hat H t]\ket{\psi}$. (3) Create a hole on site $j$ of the right system 
\begin{equation}
      \ket{\psi_l(t)} \otimes \ket{\psi_r(t)} \equiv e^{-i \hat H t}c_i^\nag \ket{\psi} \otimes c_j e^{-i \hat H t}\ket{\psi}.                                                                                                                                                                                                                                                                                                                                                                                                                                                                                                                                                                                                                                                                                                                                                                                                                                                                                                                                                                                                                                                                                                                                                                                                                                                                                                                                                                                                                                                                                                                                                                                  
\end{equation}
We abbreviate the sequence of the operations (1--3) as $\hat O(i,j)$ and illustrate the corresponding quantum circuit in the bottom of the blue box in \figc{fig:protocol}{a}. (4) Finally, measure the swap operator $\hat{\mathcal{V}}$, which interchanges the particles between the left and the right subsystem 
\begin{align}                                                                                                                                                                                                                                                                                                                                                                                                                                                                                                                                                                                                                                                                                                                                                                                                                                                                                                                                                                                                                                                                                                                                                                                                                                                                                                                              \langle \hat{\mathcal{V}} \rangle &= \tr  [  \ket{\psi_r(t)}\bra{\psi_l(t)} \otimes \ket{\psi_l(t)}\bra{\psi_r(t)}] \nonumber\\
&= |\langle \psi_r(t) | \psi_l(t) \rangle|^2 =   |\langle c_j^\dag(t) c_i^\nag \rangle|^2 = \mathcal{G}^\text{gl}_{ij}(t) .                                                                                                                                                                                                                                                                                                                                                                                                                                                                                                                                                                                                                                                                                                                                                                                                                                                                                                                                                                               
\label{eq:swap}
\end{align}
The expectation value of the swap operator is experimentally determined by a global 50\%-50\% beam splitter operation, which is realized by tunnel-coupling the left and the right system, followed by a measurement of the parity-projected particle number~\cite{moura_alves_multipartite_2004, daley_measuring_2012, islam_measuring_2015}.

OTO correlation functions are measured in a similar fashion, \fig{fig:protocol}{b}. To begin with, we recycle the first three steps of the Green's function protocol, compiled in $\hat O(i,j)$. As a second step, the sign of the Hamiltonian needs to be inverted globally. The sign of the interaction can be flipped by ramping the magnetic field across a Feshbach resonance, as demonstrated experimentally, for instance, in the realization of negative temperature states~\cite{braun_negative_2013}. Furthermore,  by appropriately tuning the drive frequency of a modulated optical lattice, the sign of the hopping matrix element can be flipped~\cite{lignier_dynamical_2007}. Combining these already established experimental techniques, the global sign of the Hamiltonian is inverted. As a next step, the operations $\hat O(j,i)$ are applied again, leading to the time evolved state
\begin{equation}
 \ket{\psi_l(t)} \otimes \ket{\psi_r(t)} \equiv e^{i \hat H t} c_j e^{-i \hat H t}c_i^\nag \ket{\psi} \otimes c_i e^{i \hat H t}c_j e^{-i \hat H t}\ket{\psi}.
\end{equation}
The square modulus of the OTO correlators is then obtained by measuring the wavefunction overlap of the left and the right system using beam splitters as discussed before. 

For the measurement of both the Green's function and the OTO correlators, the initial state $\ket{\psi}$ can be an arbitrary pure state, such as the ground state, or a simple product state. An effective finite temperature state can be obtained for quenches from initial pure states to some final Hamiltonian. In a thermalizing system~\cite{deutsch_quantum_1991, srednicki_chaos_1994, rigol_thermalization_2008}, the effective temperature is then determined by the energy-density produced by the quantum quench.
In the case of a thermal initial state, after the first three steps of our protocol, blue box in \fig{fig:protocol}, the system is prepared in the state $\rho_l(t) \otimes \rho_r(t)$, where $\rho_\alpha(t)$ is a generic density matrix. The measurement of the swap operator $\hat{\mathcal{V}}$ yields~\cite{daley_measuring_2012}
\begin{allowdisplaybreaks}
 \begin{align}
 \langle \hat{\mathcal{V}} \rangle &= \tr \hat{\mathcal{V}} \rho_l(t) \otimes \rho_r(t) \nonumber\\ 
 &= \tr \hat{\mathcal{V}} \sum_{\mu \nu} \rho_{l,\mu}(t) \rho_{r,\nu}(t) \ket{\mu}\bra{\mu}\otimes \ket{\nu}\bra{\nu} \nonumber\\ 
 &= \tr  \sum_{\mu \nu} \rho_{l,\mu}(t) \rho_{r,\nu}(t) \ket{\nu}\bra{\mu}\otimes \ket{\mu}\bra{\nu} \nonumber \\ &= \sum_\mu \rho_{l,\mu}(t) \rho_{r,\mu}(t) = \tr \rho_{l}(t) \rho_{r}(t).
\end{align}
\end{allowdisplaybreaks}
For pure states, $\rho_{l,r}(t) = \ket{\psi_{l,r}(t)} \bra{\psi_{l,r}(t)}$, we directly obtain \eq{eq:swap}. However, at finite temperature, the measurement does not directly yield the square of the correlation function. In particular, we obtain for the Green's function protocol
\begin{equation}
 \langle \hat{\mathcal{V}}\rangle = \sum_{\mu \nu}\rho_\mu \rho_\nu  \bra{\mu} c_i^\dag c_j^\nag(t) \ket{\nu}\bra{\nu} c_j^\dag(t) c_i^\nag \ket{\mu} .
 \label{eq:VT}
\end{equation}
By contrast, the desired modulus square of the thermal Green's function would be
\begin{equation}
 |\langle c_i(t)^\dag c_j\rangle|^2 = \sum_{\mu \nu} \rho_\mu \rho_\nu  \bra{\mu} c_i^\dag c_j^\nag(t) \ket{\mu}\bra{\nu} c_j^\dag(t) c_i^\nag \ket{\nu}.
\end{equation}
Hence, at high temperatures, \eq{eq:VT} is suppressed by a factor $1/Z$, where $Z$ is the partition sum, and thus vanishes in the thermodynamic limit. A similar reasoning applies in the case of OTO correlators.

\subsection{Local Many-Body Interferometry \label{sec:local}}

Interfering two system copies globally requires beam splitter operations with high fidelity, as in each measurement for systems of size $L$ the same number of beam splitter operations have to be applied. To overcome this challenge, we introduce an alternative protocol that is scalable since it only requires two beam splitter operations irrespective of the system size.

A local beam splitter operation on site $l$ is realized by coupling the left and the right copy of the quantum system by a tunneling Hamiltonian
\begin{equation}
 \hat H_l^\text{BS} = - J_\text{BS} (a_l^\dag b_l^\nag + b_l^\dag a_l^\nag),
 \label{eq:hbs}
\end{equation}
where $a_l^\dag$ ($b_l^\dag$) creates a particle in the left (right) system. The unitary evolution under \eq{eq:hbs}, $\text{BS}_l(\tau) = \exp[-i \hat H_l^\text{BS} \tau ]$, for time $\tau_\text{BS}=\pi/4 J_\text{BS}$ defines a 50\%-50\% beam splitter operation 
\begin{equation}
 \left( \begin{array}{c}
         \bar a_{l} \\ \bar b_{l}
        \end{array}\right)  =
        \underbrace{\frac{1}{\sqrt{2}}\left(\begin{array}{cc}
         1 & -i \\
         -i & 1
        \end{array}\right)}_{=\text{BS}_l}
	 \left(\begin{array}{c}
         a_{l} \\ b_{l}
        \end{array}   
 \right).
\end{equation}
Furthermore, the phase of the beam splitter can be adjusted by applying a field gradient between the left and the right system $\hat H_l^\text{F} = \frac{h}{2} (b_l^\dag b_l-a_l^\dag a_l)$ for a duration $\tau_f$, $R(\tau_f)=\exp[-i \hat H_l^\text{F} \tau_f]$:
\begin{align}
         \overline{\text{BS}}_l=R^\dag(\phi)\, \text{BS}_l\, R(\phi)= 
        \frac{1}{\sqrt{2}}\left(\begin{array}{cc}
         1 & -i e^{-i \phi} \\
         -i e^{i \phi} & 1
        \end{array}\right),
 \label{eq:BS}
\end{align}
where $\phi=h\tau_f$. 

\begin{figure*}
  \centering 
  \includegraphics[width=.7\textwidth]{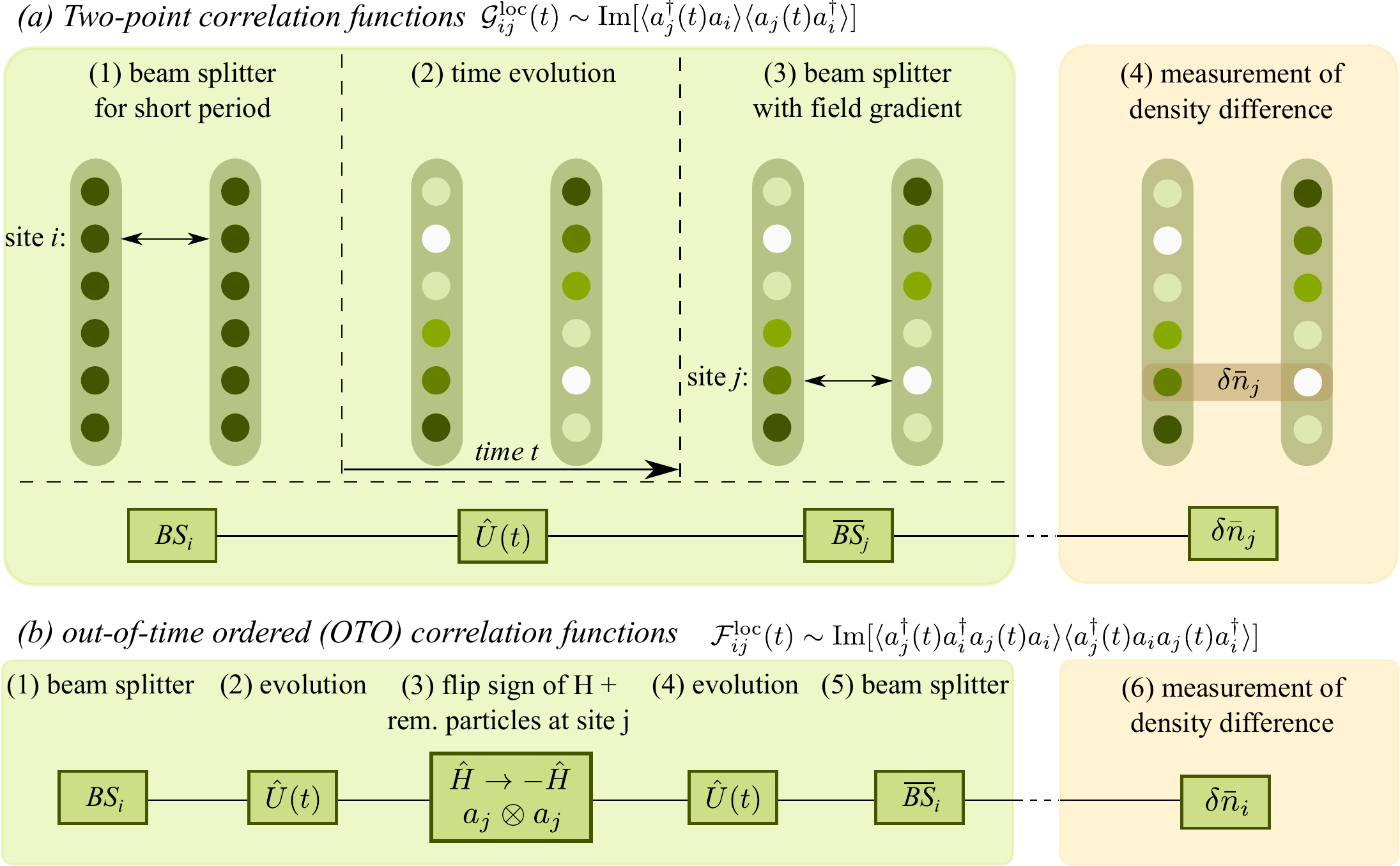}
  \caption{ \textbf{Probing dynamical correlation functions through the \emph{local} interference of two many-body states.} Using local beam-splitter operations only, our protocol measures \fc{a} time ordered correlator  $\mathcal{G}^\text{loc}_{ij}(t) \sim \im [ \langle a_j^\dag(t) a_i \rangle \langle a_j(t) a_i^\dag \rangle ]$ as well as \fc{b} the OTO correlator $\mathcal{F}^\text{loc}_{ij}(t) \sim \im [ \langle a_j^\dag(t) a_i^\dag a_j(t) a_i \rangle \langle a_j^\dag(t) a_i a_j(t) a_i^\dag \rangle ]$. A detailed description of the protocl is givne in the text.   } 
  \label{fig:protocol_local}
\end{figure*}
\begin{figure}
  \centering 
  \includegraphics[width=.48\textwidth]{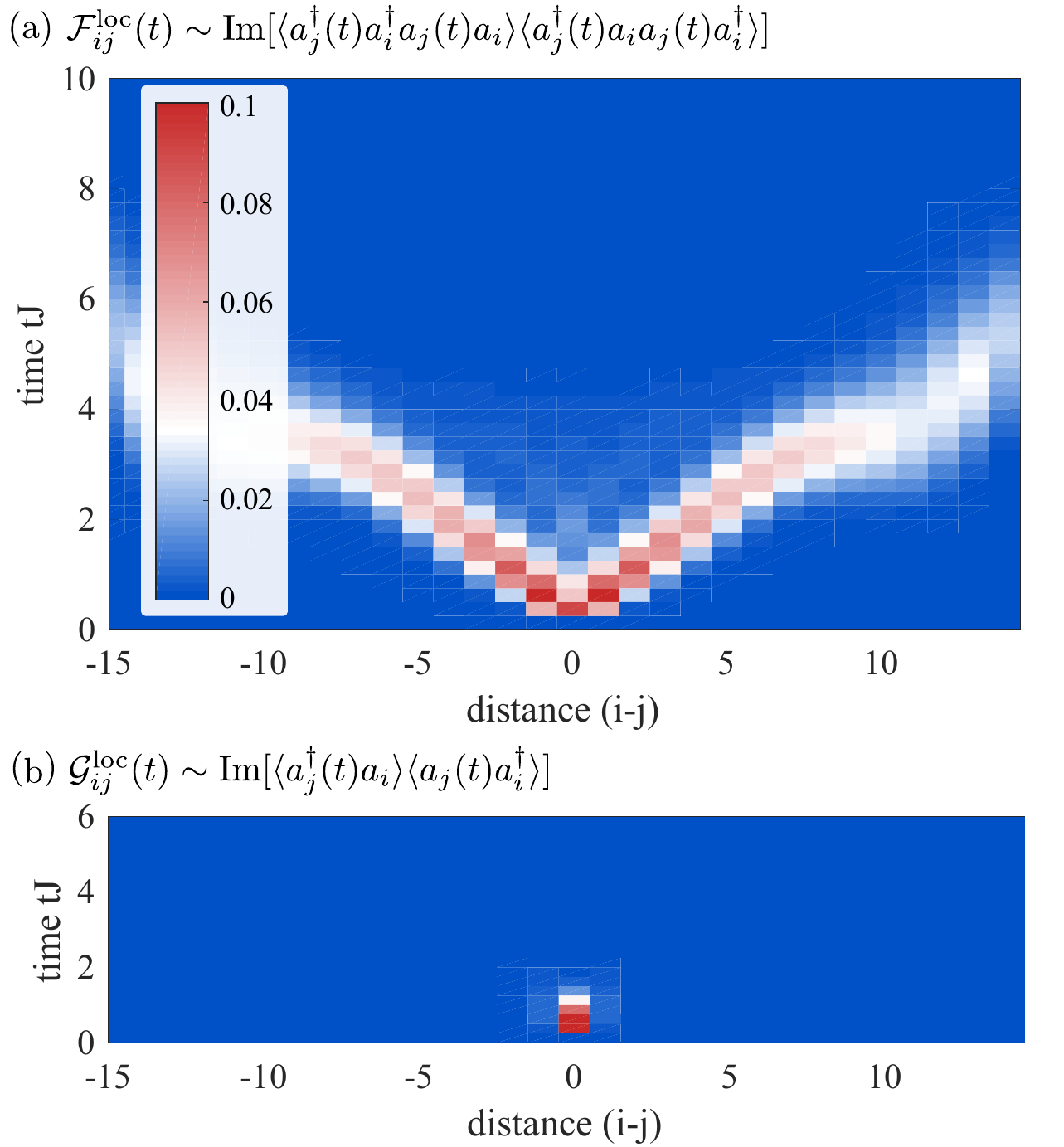}
  \caption{ \textbf{Correlation functions obtained from the local interference of two quantum states.} \fc{a} OTO correlation functions $\mathcal{F}^\text{loc}_{ij}(t)$ and \fc{b} time-ordered correlation functions $\mathcal{G}^\text{loc}_{ij}(t)$  as measured by the protocol based on local beam splitter operations (Sec.~\ref{sec:local}) contain similar information as the originally introduced correlators. The data is shown for temperature $T=4J$, interactions $U=J$, chemical potential $\mu=0$, and systems of size $L=30$ and can thus be compared to \fig{fig:otoc}.    } 
  \label{fig:otoc_loc}
\end{figure}

The time ordered Green's function for a system prepared in an arbitrary density matrix can be measured by the following sequence (\fig{fig:protocol_local}): (1) apply a beam splitter operation on site $i$ for a short duration $\tau J_\text{BS} \ll 1$. In that limit, the unitary evolution can be linearized $\text{BS}_l(\tau)=1-i \hat H_l^\text{BS} \tau+O(J_\text{BS}^2\tau^2)$. (2) Let the two copies evolve for the physical time $t$. (3) Apply a 50\%-50\% beam splitter operation on site $j$ with a phase that is detuned from the first one by $\phi=\pi/2$. (4) Finally, the density difference $\delta \bar n_j$ between the right and the left subsystem is measured. This leads to the following measurement outcome
\begin{equation}
 \mathcal{G}_{ij}^\text{loc}(t)=\langle \text{BS}^\dag_i(\tau) e^{i\hat H t} \overline{\text{BS}}_j^\dag \delta \bar n_j \overline{\text{BS}}_j e^{-i\hat H t}\text{BS}_i(\tau)\rangle.
\end{equation}
We first calculate the densities after the beam splitter operation $\overline{\text{BS}}_j$, which gives
\begin{subequations}
\begin{align}
\bar n_{l,j}&=\bar a^\dag_{j} \bar a_{j}= \frac12 (a_j^\dag-b_j^\dag)(a_j-b_j)\\
\bar n_{r,j}&=\bar b^\dag_{j} \bar b_{j}= \frac12 (a_j^\dag+b_j^\dag)(a_j+b_j).
\end{align}
\end{subequations}
Computing the density difference between the right and the left system, we find
\begin{align}
\delta \bar n_{j}=\bar n_{l,j}-\bar n_{r,j}= -(b^\dag_{j} a^\nag_{j}+ a^\dag_{j} b^\nag_{j}).
\end{align}
Considering now that the duration of the first beam splitter operation on site $i$ is short and using the particle number conservation, we obtain 
\begin{equation}
 \mathcal{G}_{ij}^\text{loc}(t) = 4J_\text{BS}\tau \im \{ \langle a_j^\dag(t) a_i^\nag \rangle \langle a_j^\nag(t) a_i^\dag \rangle \} + O(J_\text{BS}^3 \tau^3).
 \label{eq:locBS_GF}
\end{equation}
The conventional time ordered one-body correlation function is defined as $\mathcal{G}_{ij}(t) = \langle a_j^\dag(t) a_i^\nag \rangle$. In our protocol, the imaginary part of the product of a particle and a hole correlation function is measured. However, we argue below that this observable carries related information as the time-ordered correlation function $\mathcal{G}_{ij}(t)$; see also \figc{fig:otoc_loc}{b}. 

OTO correlators are measured by a straight forward extension; \figc{fig:protocol_local}{b}: (1) Apply a beam splitter operation for a short duration $\tau J_\text{BS} \ll 1$ at site $i$. (2) Let the system evolve for a physical time $t$. (3) Use single-site addressing to remove a particle on site $j$ in both copies. (4) Flip the sign of the Hamiltonian $\hat H\to -\hat H$, as suggested in the previous section. Let the system evolve in time for the duration $t$. (5) Apply the 50\%-50\% beam splitter operation $\overline{\text{BS}}_i$ on site $i$. Evaluating these steps, we find
 \begin{align}
 &\mathcal{F}_{ij}^\text{loc}(t) = \langle \text{BS}^\dag_i(\tau) e^{i\hat H t} a_j^\dag b_j^\dag e^{-i\hat H t} \overline{\text{BS}}_i^\dag \delta \bar n_i\nonumber\\
 &\qquad\qquad\times\,\overline{\text{BS}}_i e^{i\hat H t} a_j b_j e^{-i\hat H t}\text{BS}_i(\tau)\rangle \nonumber\\
 &=4J_\text{BS}\tau \im \{ \langle a_j^\dag(t) a_i^\dag a_j^\nag(t) a_i^\nag \rangle \langle a_j^\dag(t) a_i^\nag  a_j^\nag(t) a_i^\dag \rangle \} + O(J_\text{BS}^3 \tau^3).
\end{align}
This expression corresponds to the product of two OTO correlation function. Four point correlators in spin systems can be obtained with related protocols~\cite{serbyn_interferometric_2014}.

The OTO correlator $\mathcal{F}_{ij}^\text{loc}(t)$ obtained from local interference contains at the considered temperatures essentially the same information as the one we originally introduced. As the protocol measures the imaginary part of a product of two OTO correlators, it starts out at zero. The scrambling across the quantum state manifests itself in the linear propagation of a wave-packet in $\mathcal{F}_{ij}^\text{loc}(t)$ [see \figc{fig:otoc_loc}{a}] from which light-cone and butterfly velocities can be extracted. In \figc{fig:otoc_loc}{a}, we once again attribute the superballistic spread of information, which kicks in at $|i-j|\gtrsim 7$, to the finite MPO bond dimension of 400. Similarly, $\mathcal{G}_{ij}^\text{loc}(t)$ starts off at zero but then develops a peak that quickly decays; \figc{fig:otoc_loc}{b}. From that we determine the quasiparticle lifetime $\tau J \sim 0.32$ which corresponds roughly to half the lifetime obtained for the Green's function $\mathcal{G}_{ij}(t)$. This factor can be attributed to the fact that here the product of two correlation functions is measured. 

With the protocols discussed so far, static one-body correlation functions can be measured by setting the physical time $t=0$. 
Moreover, a generalization of the local protocol makes it possible to measure static correlations functions of \emph{arbitrary} order. Specifically, correlators of $\delta \bar n_i$ determine one-body correlation functions of the original many-body state:
\begin{equation}
 \langle \delta \bar n_i \delta \bar n_j \rangle = 2\langle a_i^\dag a_j^\nag \rangle\langle a_i^\nag a_j^\dag \rangle.
\end{equation}
Here we used that the left and the right initial states are identical. Higher order static correlation functions in the creation $a^\dag_i$ and annihilation operators $a_i$ are straightforwardly obtained by measuring higher order correlators in $\delta \bar n_i$. We emphasize that this protocol scales favorable with system size, and that correlators between arbitrary sites and of arbitrary order can be taken in a single shot by performing the beam splitter operations on the full system.

The density matrix describing the quantum state of a system can be expressed as
\begin{equation}
\hat{\rho} = \mathcal{N} \sum_{i_1,...i_N} r_{i_1,...i_N} \hat{\sigma}_{i_1} \otimes ...\otimes \hat{\sigma}_{i_N},
\end{equation}
where $\mathcal{N}$ is a normalization constant and the $\hat{\sigma}_{i_j}$ constitute a suitable basis \cite{James2001}. In the case of fermions or hard-core bosons, one possible choice for the basis are the Pauli matrices. The knowledge of correlators up to sufficient order makes it possible to determine the so-called Stokes parameters $r_{i_1,...i_N}$ and thereby to reconstruct the density matrix, which paves the way for the full state tomography of quantum states with massive particles.

\subsection{Measuring Dynamical Density Correlators \label{app:dens}}

In this section we discuss two different possibilities to measure dynamical density correlation functions and thereby observe their diffusive behavior. 

The dynamic structure factor $S(k,\omega)$, which is the spatial and temporal Fourier transform of the density correlator $C_n(x,t)$, can be measured with Bragg spectroscopy~\cite{clement_2009, ernst_probing_2010}. In Bragg spectroscopy, the detuning of the two laser beams sets the frequency $\omega$ and the angle between the beams the transferred momentum $k$. A measurement of the absorption of the system as a function of $k$ and $\omega$ directly maps out the dynamic structure factor $S(k,\omega)$. Diffusion manifests itself in the wavevector and frequency resolved structure factor $S(k,\omega)$ as Lorentzian peaks with half-width-half-maximum that scales as $Dq^2$.

It is furthermore possible to measure the dissipative response $\langle [n_i(t), n_j] \rangle$ to a local perturbation of the system in a quantum gas microscope. To this end, a local potential $\delta H = n_j \delta \mu$ is created at site $j$ by applying a laser for a short time $\tau$, yielding the time evolution $\exp[-i \delta H \tau] \sim 1-i \delta H \tau + O(\delta \mu^2 \tau^2)$. Measuring the density at site $i$ after the unitary time evolution for duration $t$ we obtain
\begin{equation}
\chi_{ij}^\text{loc}(t) = \langle n_i(t) \rangle + i \delta \mu \tau  \langle [n_i(t), n_j] \rangle + O(\delta \mu^2 \tau^2).
\end{equation}
In equilibrium, the fluctuation-dissipation theorem provides an exact relation between $\langle [n_i(t), n_j] \rangle$ and $\langle n_i(t) n_j \rangle$. The accurate measurement of the former therefore enables the observation of diffusive response in the dynamical density correlator.

\end{document}

%% file: otoc_v15.bbl
%